\documentclass[twocolumn]{jpsj2}
\usepackage{graphicx, color}
\usepackage{bm}
\def\runauthor{Online-Journal Subcommittee}

\title{%
AC Susceptibility of the Dipolar Spin Ice Dy$_2$Ti$_2$O$_7$:\\
Experiments and Monte Carlo Simulations
}

\author{%
Hiroshi Takatsu$^{1,}$\thanks{E-mail address: takatsu@tmu.ac.jp}, 
Kazuki Goto$^1$, Hiromi Otsuka$^1$, Ryuji Higashinaka$^1$, Kazuyuki Matsubayashi$^2$, \\
Yoshiya Uwatoko$^2$, and Hiroaki Kadowaki$^1$ 
}

\inst{%
$^1$Department of Physics, Tokyo Metropolitan University, Hachioji-shi, Tokyo 192-0397, Japan\\
$^2$Institute for Solid State Physics, University of Tokyo, Kashiwanoha 5-1-5, Kashiwa, Chiba 277-8581, Japan
}

\recdate{\today}

\abst{%
Experimental data of the frequency-dependent ac susceptibility [$\chi_{\mathrm{ac}}(\omega)$] 
for the dipolar spin ice Dy$_2$Ti$_2$O$_7$ has been analyzed by
Monte Carlo simulations on the basis of the single-spin-flip Metropolis algorithm. 
We have directly evaluated $\chi_{\mathrm{ac}}(\omega)$ by applying an ac magnetic field.
We found that this simulation reasonably reproduces the experimental behavior of $\chi_{\mathrm{ac}}(\omega)$ 
in the temperature range from 0.6 to 1.0~K, where dilute magnetic monopoles diffusively move.
The conversion factor from simulation time to real time, i.e., 
the rate of hopping of monopoles to nearest-neighbor sites,
strongly depends on temperature. 
}

\kword{%
Dy$_2$Ti$_2$O$_7$, spin ice, ac susceptibility, Monte Carlo simulation
}

\begin{document}
\maketitle

\section{Introduction}
Geometrically frustrated spin systems have attracted much attention for the realization of novel ground states
and their exotic magnetic phenomena~\cite{Diep,LaCroix}.
Remarkable examples are spin ice materials such as 
Dy$_2$Ti$_2$O$_7$ (DTO)~\cite{RamirezNature1999,BramwellScience2001}, 
Ho$_2$Ti$_2$O$_7$~\cite{HarrisPRL1997,BramwellPRL2001}, and Ho$_2$Sn$_2$O$_7$~\cite{MatsuhiraJPCM2000,KadowakiPRB2002},
where Dy or Ho atoms occupy corners of tetrahedra.
In these compounds,
a strong crystalline electric field forces the magnetic moments of rare-earth ions 
to lie along the local $\langle111\rangle$ symmetry direction~\cite{BramwellScience2001}. 
This Ising anisotropy combined with the effective ferromagnetic nearest-neighbor-interaction $J_{\mathrm{eff}}$, 
on the order of 1--2~K,~\cite{BramwellScience2001}
stabilizes the so-called 2-in 2-out structure~\cite{HarrisPRL1997}, 
analogously to the ice rule of the proton configurations in water ice~\cite{BernalJCP1933}. 
The ground states of the spin ice are macroscopically degenerate, 
giving rise to residual entropy~\cite{RamirezNature1999}.
Recent theories pointed out that elementary excitations from such degenerate ground states 
can be viewed as magnetic point defects~\cite{RyzhkinJETP2005}, i.e., monopoles~\cite{CastelnovoNature2008}. 
A pair consisting of a monopole and an antimonopole, created by flipping a spin, 
can be fractionalized into two individual excitations 
and diffusively move by successively flipping neighboring spins. 
Therefore,
the magnetic dynamics of the spin ice is an interesting problem,
which enables us to study monopole motion and relaxation~\cite{JaubertNaturePhys2009,Jaubert2011}. 

The spin dynamics of the spin-ice materials studied by 
the ac susceptibility $\chi_{\mathrm{ac}}(\omega)$ 
shows characteristic behavior associated with the single relaxation 
time $\tau$. 
The single $\tau$ is a simple approximation and is often used to describe the dynamics of 
these materials~\cite{MatsuhiraJPCM2001,SnyderPRB2004,MatsuhiraJPSJ2011,YaraskavitchPRB2012,RevellNaturePhys2012,QuilliamPRB2011}.
Measurements of $\chi_{\rm ac}(\omega)$ on single-crystalline samples and powder samples of DTO have revealed that 
the temperature dependence of $\tau$ exhibits a plateau-like behavior between 2 and 10~K 
and a rapid increase upon cooling below 2~K~\cite{MatsuhiraJPCM2001,SnyderPRB2004,MatsuhiraJPSJ2011,YaraskavitchPRB2012,RevellNaturePhys2012}.
Monte Carlo (MC) simulations~\cite{JaubertNaturePhys2009,Jaubert2011} 
reproduced this behavior of $\tau$ at $T\ge2$~K,
supporting the picture of diffusively moving monopoles to a certain extent. 
More recent MC simulations and experiments 
showed that the rate of hopping of monopoles to nearest neighbor sites (i.e., diffusion constant)
depends strongly on temperature, giving rise to a marked increase in $\tau$ below 1~K~\cite{RevellNaturePhys2012}.

In this work,
we have performed dynamical MC simulations to obtain the $\chi_{\mathrm{ac}}(\omega)$
of the spin ice DTO by applying an ac magnetic field.
We used the single-spin-flip Metropolis algorithm for
the dipolar spin-ice model~\cite{HertogPRL2000,MelkoJPCM2004,RuffPRL2005}
in order to simulate the diffusive dynamics of monopoles~\cite{JaubertNaturePhys2009,Jaubert2011}.
The purpose of the present work is 
to directly calculate $\chi_{\mathrm{ac}}(\omega)$ and compare it with experimental data of 
$\chi_{\mathrm{ac}}(\omega)$ for DTO, and then to explore the monopole dynamics.
In particular,
our main effort is focused on investigating the {\it frequency dependence} of $\chi_{\mathrm{ac}}(\omega)$,
in the hope that the present method can be used to analyze experimental data of $\chi_{\mathrm{ac}}(\omega)$ published 
in Refs.~14-19, 23 and 24.
We compare our $\tau$ with Jaubert's $\tau$, 
which has been computed from 
the autocorrelation of magnetization~\cite{JaubertNaturePhys2009,Jaubert2011}.

\section{Calculations and Experiments}
The Hamiltonian used in the simulations is that of the dipolar spin ice model~\cite{HertogPRL2000,MelkoJPCM2004,RuffPRL2005}:
\begin{align}
\mathcal{H} = &-\mu_{\mathrm{eff}}\sum_{i,a}\bm{S}_i^a\cdot\bm{H} 
               - \sum_{\langle(i,a),(i,b)\rangle}J_{i,a;j,b} \bm{S}_i^a\cdot\bm{S}_j^b \notag \\
              &+Dr_{nn}^3\sum_{i,j,a,b}\biggl[\frac{\bm{S}_i^a\bm{S}_j^b}{|\bm{R}_{ij}^{ab}|^3}-\frac{3(\bm{S}_i^a\cdot \bm{R}_{ij}^{ab})(\bm{S}_j^b\cdot \bm{R}_{ij}^{ab})}{|\bm{R}_{ij}^{ab}|^5} \biggr],
\end{align}
where $\bm{S}_i^a$, with the unit length $|\bm{S}_i^a| = 1$, 
represents the spin vector parallel to the local $\langle111\rangle$ direction
at the sublattice site $a$ in the unit cell of the fcc lattice site $i$.
The first term represents the Zeeman interaction between the spins with 
the effective moment $\mu_{\mathrm{eff}}=9.866~\mu_{\mathrm{B}}$ and the magnetic field $\bm{H}$.
The second and third terms are the exchange and dipolar interactions, respectively.
We used the nearest-, second-, and third-neighbor interactions of
$J_{1} = -3.41$~K, $J_{2} = 0.14$~K, and $J_{3} = -0.025$~K, respectively,
and the dipolar interaction parameter $D=1.32$~K~\cite{YavorsPRL2008,TabataPRL2006}.
These parameters were adjusted to reproduce many experimental data of DTO,
improving the simulation~\cite{HertogPRL2000} using the two parameters $J_{1}$ and $D$. 
We also performed simulations with $J_{1} = -3.72$~K and $D=1.41$~K ($J_2=J_3=0$),
in order to compare our results with previous works in Refs.~12 and 13. 

To obtain $\chi_{\rm ac}(\omega)$, we performed MC simulations on the basis of
the single-spin-flip Metropolis algorithm. 
In this simulation, the MC-step per spin, $t_{\rm MC}$, is used also 
as the time of an oscillating ac magnetic field $\bm{H}(t_{\rm MC}) = \bm{H}_0\sin(\omega t_{\rm MC})$,
where $\omega=2\pi f$.
The MC-step $t_{\rm MC}$ is regarded as very similar to the real time $t_{\rm real}$ of the measurements of $\chi_{\rm ac}(\omega)$,
that is, we assume $t_{\rm MC} = A t_{\rm real}$,
whose details will be explained later.
The real and imaginary parts of $\chi_{\mathrm{ac}}(\omega) [= \chi'(\omega) -i \chi''(\omega)]$
were calculated as
\begin{align}
&\chi' (\omega){\bm{H}}_0=\frac{1}{N_{\rm MC}}
\sum_{t_{\rm MC}=1}^{N_{\rm MC}}{\bm{M}}(t_{\rm MC})\sin(\omega t_{\rm MC}),\cr
&\chi''(\omega){\bm{H}}_0=\frac{1}{N_{\rm MC}}
\sum_{t_{\rm MC}=1}^{N_{\rm MC}}{\bm{M}}(t_{\rm MC})\cos(\omega t_{\rm MC}),
\end{align}
where $N_{\rm MC}$ is the number of total MC steps 
and 
${\bm{M}}(t_{\rm MC})$ represents the uniform magnetization sampled at each $t_{\rm MC} =1,2,\cdots N_{\rm MC}$. 
Note that a similar method has been employed for calculating
the $\chi_{\mathrm{ac}}(\omega)$ of spin glass~\cite{PiccoPRB2001}.

To compare the simulated $\chi'(\omega)$ and $\chi''(\omega)$ with the experimental values,
the $T$-dependent conversion factor from the MC time to the real time, $A(T) \equiv  t_{\rm MC}/t_{\rm real}$, 
was used.
This factor $A(T)$ was estimated by comparing the $\chi_{\rm ac}(\omega)$ of 
simulations and experiments.
We simulated the systems with a system size of 2916 spins up to $N_{\rm MC}=4\times10^5$~MC steps/spin. 
A periodic boundary condition was used.
We typically used a magnetic field strength of $|{\bm{H}}_0|=100$~Oe.
Although this value is larger than the experimentally used $|{\bm{H}}_0|$~\cite{MatsuhiraJPSJ2011,YaraskavitchPRB2012,RevellNaturePhys2012,MatsuhiraJPCM2001,SnyderPRB2004,SnyderNature2001,BovoNatureComm2013,QuilliamPRB2011},
we confirmed that $\chi'(\omega = 0)$ does not depend on $|{\bm{H}}_0|$ for
$|{\bm{H}}_0|=20$, 50, 100, or 200~Oe at 0.6 or 1~K.
We also confirmed that $\chi_{\mathrm{ac}}(\omega)$ is isotropic with respect to the field direction,
which was tested by applying a magnetic field along both $[111]$ and $[11\bar{2}]$.
This result is in fact consistent with the experimental observations~\cite{MatsuhiraJPSJ2011,YaraskavitchPRB2012}.

It should be noted that a temperature-independent factor $A(T)$ was used in studies described 
in Refs.~12 and 13 to compare between experimentally evaluated $\tau$ 
and calculated $\tau$ by MC simulations, where the autocorrelation of magnetization was used~\cite{JaubertNaturePhys2009,Jaubert2011}.
It was argued that the $T$-independent factor $A(T)$ means that all the monopole or spin dynamics 
are taken into account by the MC simulation,
which uses the Metropolis algorithm with a single-spin-flip dynamics, where monopoles move diffusively.
In contrast, we need to use $T$-dependent $A(T)$, 
implying that there are some other mechanisms not taken into account by the MC simulation.
Roughly speaking, 
the values of $A(T)$ were evaluated from peak positions of $\chi''(\omega)$:
$A(T) = f_{\rm p, exp}/f_{\rm p, MC}$, where $f_{\rm p, MC}$ and $f_{\rm p, exp}$ are 
the peak positions of $\chi''(\omega)$ in the MC simulations and experiments, respectively.
These peak positions are related to $\tau$ for the single-$\tau$ approximation:
$\tau_{\rm MC} = 1/2\pi f_{\rm p, MC}$ and $\tau_{\rm exp} = 1/2\pi f_{\rm p, exp}$.
A similar method for estimating such a conversion factor 
has been used in a recent work~\cite{RevellNaturePhys2012}.

We performed measurements of $\chi_{\mathrm{ac}}(\omega)$ on 
a single-crystalline sample of DTO, 
because there is a slight discrepancy in previous experimental results between single crystals and polycrystals at about 2~K (Fig.~\ref{fig.3}).
We used a rectangular crystal
cut into a thin plate with a size of approximately $3\times5\times0.2$~mm$^3$ and a mass of 19.6~mg.
A wide plane of the sample includes the [$111$] and [$11\bar{2}$] directions.
Because the demagnetization factors for these directions are small enough ($N = 0.08$ and 0.05, respectively~\cite{Aharoni1998}),
we did not perform the demagnetization correction.
Note that, as will be discussed later,
the demagnetization effect sensitively affects the precise estimation of $\tau_{\rm exp}$ 
for spin ice compounds~\cite{QuilliamPRB2011,YaraskavitchPRB2012}.
$\chi_{\mathrm{ac}}(\omega)$ was measured using a commercial 
superconducting quantum interference device (SQUID) magnetometer (Quantum Design MPMS) 
down to 1.8~K and a mutual inductance method below 2~K in a $^3$He refrigerator.
An ac magnetic field $H_{\mathrm{ac}}$ with a strength of 0.1, 1, or 5~Oe-rms 
was applied along the [$11\bar{2}$] direction.
$\chi_{\mathrm{ac}}(\omega)$ at different frequencies ranging from 1~Hz to 1.5~kHz was measured 
and the background signal was subtracted. 

\section{Results}
\subsection{Temperature dependence of $\chi_{\mathrm{ac}}(\omega)$ in the dc limit}
\begin{figure}[t]
\begin{center}
 \includegraphics[width=0.45\textwidth]{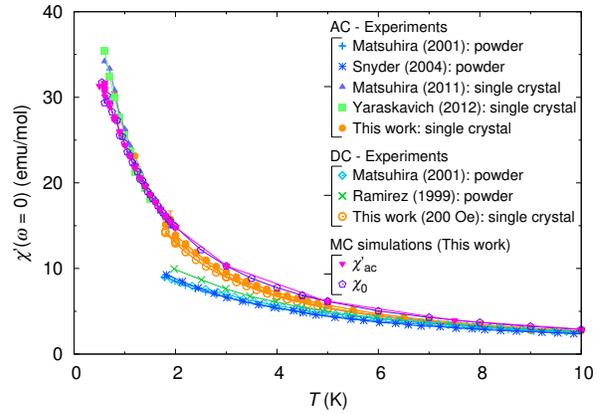}
\caption{
(Color online)
Temperature dependence of the real part of $\chi_{\rm ac}(\omega)$
in the dc limit, $\chi'(\omega = 0)$ and dc susceptibility.
The simulated $\chi'(\omega = 0)$ well reproduces the
experimental results obtained for single crystals, but not those for powder samples. 
It is also in agreement with the standard zero-field susceptibility $\chi_{0}\propto{\langle M^2\rangle}/{T}$ 
calculated by simulations.
}
\label{fig.3}
\end{center}
\end{figure}
In Fig.~\ref{fig.3},
we present the temperature dependence of the calculated and observed $\chi'(\omega)$ values, 
compared with previous results.
The dc limit of $\chi'(\omega)$ [$= \chi'(\omega=0)$] of the simulation is
in reasonable agreement with the experimental results of the single crystals. 
It also exhibits the same behavior of the zero-field susceptibility $\chi_{0}\propto{\langle M^2\rangle}/{T}$ calculated by simulations,
where $\langle M^2\rangle$ represents the fluctuation of the uniform magnetization.
We confirmed that the $\chi'(\omega=0)$ of our single crystal 
matches previous results of single crystals~\cite{MatsuhiraJPSJ2011,YaraskavitchPRB2012}, 
smoothly varying at about 2~K (Fig.~\ref{fig.3}). 
These results ensure the reliability of the results of the simulations and experiments at the first step.

\subsection{Frequency dependence of $\chi_{\mathrm{ac}}(\omega)$}
\begin{figure}[t]
\begin{center}
 \includegraphics[width=0.45\textwidth]{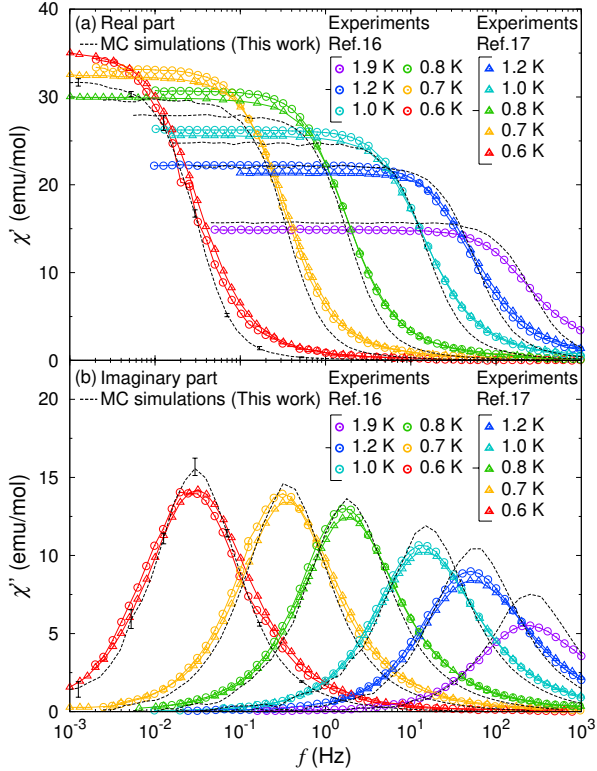}
\caption{
(Color online) 
Comparison of the frequency-dependent $\chi_{\mathrm{ac}}(\omega)$
between experiments using single crystals~\cite{MatsuhiraJPSJ2011,YaraskavitchPRB2012} and MC simulations.
(a) Real parts of $\chi_{\mathrm{ac}}(\omega)$, $\chi'(\omega)$, at several temperatures.
(b) Imaginary parts of $\chi_{\mathrm{ac}}(\omega)$, $\chi''(\omega)$, at the same temperatures
as those in (a).
Dashed lines represent the results of simulations, where
the timescale was adjusted to that of experiments by the conversion factor of $A(T)$ (see text for details).
Bars for simulated data represent error bars.
}
\label{fig.1}
\end{center}
\end{figure}
\begin{figure}[t]
\begin{center}
 \includegraphics[width=0.45\textwidth]{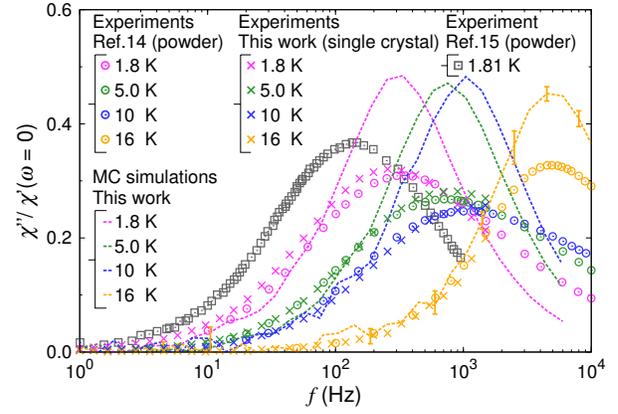}
\caption{
(Color online) 
Comparison of $\chi''(\omega)$ between experiments and MC simulations at temperatures above 1.8~K.
The experimental data in Ref.~14 obtained using a powder sample is approximately the same as
our experimental data on the single-crystalline sample obtained at the same temperatures (cross symbols). 
In contrast, there is a clear discrepancy between the data in Ref.~15
and our single-crystal data. This discrepancy is likely due to demagnetization effects~\cite{QuilliamPRB2011,YaraskavitchPRB2012}.
Dashed lines with colors represent results of MC simulations.
}
\label{fig.4}
\end{center}
\end{figure}

Figure~\ref{fig.1} shows the frequency dependence of $\chi_{\mathrm{ac}}(\omega)$.
We confirm that the calculated $\chi_{\mathrm{ac}}(\omega)$ is in reasonable agreement with 
the observations in the temperature range from 0.6 to 1.0~K,
where the spin-ice correlations strongly develop and 
the monopole picture provides a good description for DTO~\cite{RamirezNature1999,BramwellScience2001,CastelnovoNature2008}.
We can thus expect that our simulations will reproduce well 
the monopole dynamics in $\chi_{\mathrm{ac}}(\omega)$.
One can see that there is a slight discrepancy above 1.2~K,
which becomes more obvious in the behavior of $\chi''(\omega)$.
In Fig.~\ref{fig.4}, 
we present a comparison between the calculated and observed frequency dependences of 
$\chi''(\omega)/\chi'(\omega=0)$ at higher temperatures above 1.8~K.
The discrepancy between simulations and observations becomes clearer at higher temperatures.
We will discuss this issue in Sect.~4.

\begin{figure}[h]
\begin{center}
 \includegraphics[width=0.45\textwidth]{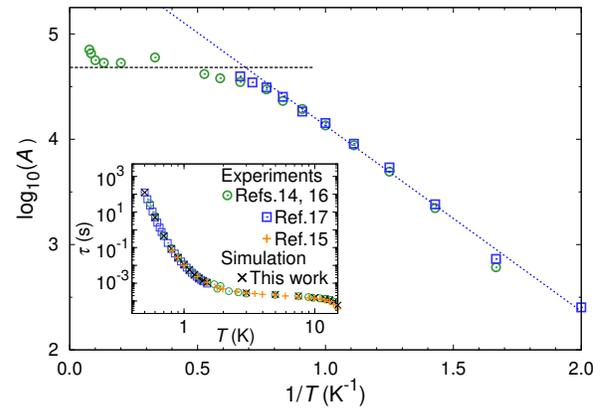}
\caption{
(Color online) 
The conversion factor $A(T)$ from real time to MC simulation time
is plotted as a function of $1/T$.
The inset shows the temperature dependence of the relaxation time $\tau$.
Cross symbols in the inset indicate the $\tau$ of our simulations,
which were adjusted by $A(T)$. 
According to Ref.~18, $\tau$ in the graph was rescaled by multiplying 
$\tau$ in Refs. 14-17 by a factor of $1/2\pi$.
Moreover, $\tau$ in Ref.~15 is multiplied by a factor of 0.5 in addition to $1/2\pi$, 
because demagnetization effects cause a clear discrepancy of the data in 
Ref.~15, as shown in Fig.~\ref{fig.4}~\cite{SnyderPRB2004,YaraskavitchPRB2012}.
}
\label{fig.5}
\end{center}
\end{figure}
The temperature dependence of the conversion factor $A(T)$,
estimated using the experimental data in Refs.~14, 16, and 17, is plotted in Fig.~\ref{fig.5}.
$A(T)$ shows a crossover behavior; it is constant between 2 and 10~K and rapidly changes upon cooling.
In particular, it exhibits the Arrhenius-type temperature dependence $A(T)\propto \exp(-\varDelta/T)$ below 1~K.
The activation energy $\varDelta$ is roughly obtained as $\varDelta=4$~K, 
which is consistent with the previous result~\cite{RevellNaturePhys2012}.
These imply that the dynamics that is not taken into account by the MC simulation
becomes crucial for the anomalous increase in $\tau_{\rm exp}$ below 1~K.
In the Fig.~\ref{fig.5} inset, $\tau_{\rm MC}$ scaled by $A(T)$ and 
$\tau_{\rm exp}$ evaluated from several works are summarized.
Note that the data in Ref.~15 is adjusted by multiplying it by a factor of 0.5
since demagnetization effects cause a definite discrepancy of the data of
Ref.~15, as also shown in Fig.~\ref{fig.4}~\cite{SnyderPRB2004,YaraskavitchPRB2012}.
It is known that large demagnetization effects affect the precise estimation of 
$f_{\rm p, exp}(=1/2\pi \tau_{\rm exp})$~\cite{QuilliamPRB2011,YaraskavitchPRB2012}.

\subsection{Temperature dependence of $\tau_{\rm MC}$}
\begin{figure}[t]
\begin{center}
 \includegraphics[width=0.45\textwidth]{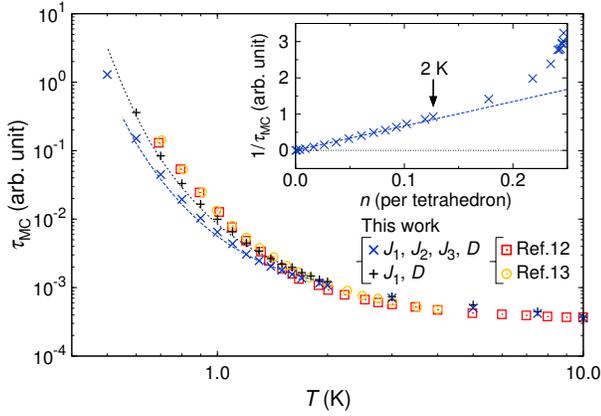}
\caption{
(Color online) 
Temperature dependence of relaxation time in MC simulations, $\tau_{\mathrm{MC}}$.
Results under parameters of $J_{1}$-$J_{2}$-$J_{3}$-$D$ and of only $J_{1}$-$D$ 
slightly deviate from each other. This is significant at temperatures below 2~K.
Instead,
both $\tau_{\mathrm{MC}}$ values show the same temperature dependence of 
the inverse of the monopole density $n$:
the dashed line shows the result of the $J_{1}$-$J_{2}$-$J_{3}$-$D$ calculation, 
and the dotted line shows the result of the $J_{1}$-$D$ calculation.
In order to compare our results with the results in Refs.~12 and 13,
the $y$-axis of $\tau_{\mathrm{MC}}$ is shifted so that it matches 
the value of Ref.~13 at 2~K. 
The same constant is used for shifting $\tau_{\mathrm{MC}}$ for both results of 
the $J_{1}$-$J_{2}$-$J_{3}$-$D$ and $J_{1}$-$D$ calculations.
The inset shows $1/\tau_{\mathrm{MC}}$ as a function of $n$ per tetrahedron.
The linear relation is realized at temperatures below about 2~K.
}
\label{fig.2}
\end{center}
\end{figure}

In Fig.~\ref{fig.2},
we present $\tau_{\rm MC}$ as a function of temperature,
which is estimated from the peak positions of $\chi''(\omega)$ using the relation 
$\tau_{\mathrm{MC}} = 1/2\pi f_{\rm{p,MC}}$. 
We found that $\tau_{\mathrm{MC}}$ continuously changes upon cooling, and
approximately varies with the inverse of the monopole density $n$,
which is calculated from the number of 3-in 1-out spin configurations per tetrahedron. 
As is clear in the Fig.~\ref{fig.2} inset, one can see a linear relation between $1/\tau_{\mathrm{MC}}$ and $n$ 
at temperatures below about 2~K where the monopole density $n$ is small. 
Note that 
such a linear relation is theoretically predicted 
for the single-spin-flip Metropolis dynamics~\cite{CastelnovoPRB2011,RyzhkinJETP2005}.

We found that the $\tau_{\mathrm{MC}}$ values are slightly different 
between the simulations using the parameters of $J_{1}$-$J_{2}$-$J_{3}$-$D$ and of only $J_{1}$-$D$.
This discrepancy becomes apparent at temperatures below 2~K
and suggests that additional parameters of $J_{2}$ and $J_{3}$ 
give a more quantitative behavior of $\tau_{\mathrm{MC}}$.
In fact, it is known that
these parameters are important for describing 
the ground state in a magnetic field~\cite{TabataPRL2006,RuffPRL2005}.
Nevertheless, we note that the change in $A(T)$ 
is much larger at $T<2$~K [Fig.~\ref{fig.5}].
This result
implies that the other mechanism 
is more crucial than the difference between these additional exchange parameters
and that it plays an important role in the rapid increase in $\tau_{\rm exp}$ below 1~K.

We confirmed that 
the temperature dependence of $\tau_{\rm MC}$ is in reasonable agreement with 
the results in Refs.~12 and 13.
The slight discrepancy is  probably due to the effect of the estimation of $\tau_{\mathrm{MC}}$:
it was extracted by fitting multiple relaxation times in the autocorrelation function~\cite{JaubertNaturePhys2009,Jaubert2011},
whereas $\tau_{\mathrm{MC}}$ was evaluated from the peak positions of $\chi''(\omega)$
as a single-$\tau$ in our simulations [Fig.~\ref{fig.1}(b)].
This discrepancy is also negligible for the change in $A(T)$,
and therefore we expect that both $\tau_{\mathrm{MC}}$ extracted from $\chi''(\omega)$ 
and from the autocorrelation function can be useful for the estimation of $\tau$ 
in simulations.

\section{Discussion}
The frequency dependence of $\chi_{\mathrm{ac}}(\omega)$ of DTO was analyzed by MC simulations
based on the single-spin-flip Metropolis algorithm.
It is realized that the observed $\chi_{\mathrm{ac}}(\omega)$ is 
adequately characterized by simulated behaviors,
particularly below 1~K, where the spin-ice correlation develops.
This means that 
the diffusive motion of monopole defects dominates the dynamics,
giving rise to a nearly single-$\tau$ frequency dependence of $\chi_{\mathrm{ac}}(\omega)$.
In contrast,
$\chi_{\mathrm{ac}}(\omega)$ at higher temperatures deviates from that in the simulations,
which is apparent in the temperature range of 2--10~K.
This result is likely due to 
the dynamics being not fully simulated by our numerical procedure.
It is considered that 
in an actual material, other complex motions can appear at $T>2~$K, which cause 
the distribution of $\tau_{\rm exp}$ and spread the frequency dependence of $\chi_{\mathrm{ac}}(\omega)$. 
In fact, the density of all-in and all-out double defects tends to increase above 4~K~\cite{Jaubert2011}.
Such contributions may lead to additional effects on and interactions with diffusively moving monopole defects,
which affect relaxation processes and time. 
Further improvement in the estimation of $\tau$
beyond the single-$\tau$ model is needed in this temperature region.

We confirmed that $A(T)$ is temperature-dependent. 
In particular, such a $T$-dependent $A(T)$ appears at temperatures below 1~K.
This result suggests that the diffusive motion of monopoles depends on temperature, 
i.e., the diffusion constant is temperature-dependent. 
It is reminiscent of the fact that a strong field associated with spin-ice correlations 
affects the flipping of a spin, or impurities confine the motion of monopoles,
which result in a temperature-dependent hopping rate of monopoles~\cite{RevellNaturePhys2012}.
This is characterized by the Arrhenius-type behavior below 1~K [Fig.~\ref{fig.5} inset].
Note that such type of diffusion constant has recently been found in a vortex dynamics 
in superconducting films, which show an exponential decay of the diffusion constant,
possibly originating from the pinning of vortex cores~\cite{GasparovPRB2012}.

We realized that $A(T)$ in our simulations exhibits approximately 
the same behavior as the conversion factor given in Ref.~18, 
in which 
$\tau_{\rm MC}$ was also computed by MC simulations on the basis of the single-spin-flip dynamics,
while considering open boundary conditions and the surface and impurity effects~\cite{RevellNaturePhys2012}.
In our simulations,
the periodic boundary condition was used, 
and the surface and impurity effects were not taken into account.
We thus think that, within the single-spin-flip dynamics and our simulation accuracies,
the qualitative behavior of $A(T)$ can be captured.
However, such conditions and effects are important for describing the detailed behavior of 
$\chi_{\rm ac}(\omega)$ in the dc limit~\cite{RevellNaturePhys2012}.
Finally, it is also worth noting here that
the ac magnetic response in simulations matching the experimental behavior 
below 1~K implies that the motion of magnetic monopoles in the spin ice
originates from diffusion, as suggested in several works~\cite{CastelnovoNature2008,JaubertNaturePhys2009,MorrisScience2009,CastelnovoPRB2011}, since the single-spin-flip dynamics simulates the diffusive motion of monopoles or spins.
It is reminiscent of a nearly single-$\tau$ curve of 
the frequency-dependent relative permittivity related to
the electron-gas-like behavior of $\mathrm{H}_3\mathrm{O}^+$ and $\mathrm{OH}^-$ ionic defects in water ice~\cite{Vivtor}, 
analogous to the behavior of the monopole and antimonopole simulated in our MC simulations.
Therefore, 
a hallmark of the diffusive motion of monopoles
in $\chi_{\rm ac}(\omega)$ can be seen in 
the simulated curve of $\chi''(\omega)$, which 
reasonably reproduces the experimental results.
\section{Conclusions}
We have investigated the frequency dependence of 
the ac susceptibility $\chi_{\mathrm{ac}}(\omega)$ of Dy$_2$Ti$_2$O$_7$ by Monte Carlo (MC) simulations
on the basis of a single-spin-flip Metropolis algorithm, where an ac magnetic field is applied 
and the ac magnetization and $\chi_{\mathrm{ac}}(\omega)$ are computed.
We demonstrated that the calculated $\chi_{\mathrm{ac}}(\omega)$ well reproduces 
the experimentally measured $\chi_{\mathrm{ac}}(\omega)$ in the temperature range from 0.6 to 1.0~K, 
where spin-ice correlations develop.
We consider the $T$-dependent conversion factor of simulation time to real time,
in order to reproduce the marked increase in relaxation time below 1~K.
This means that there are some mechanisms for the dynamics that are not taken into account in the simulations,
suggesting that
the diffusion constant or the rate of hopping of monopoles to nearest-neighbor sites strongly depend on temperature.
The present study suggests that our method of directly calculating $\chi_{\mathrm{ac}}(\omega)$ 
from the MC simulation is useful for the analyses of the experimentally measured $\chi_{\mathrm{ac}}(\omega)$
of the spin ice in the low-$T$ state~\cite{MatsuhiraJPSJ2011,YaraskavitchPRB2012,RevellNaturePhys2012,MatsuhiraJPCM2001,SnyderPRB2004,SnyderNature2001,QuilliamPRB2011} 
as well as in a dc magnetic field~\cite{BovoNatureComm2013,MatthewsPRB2012,H.TakatsuJPSJL2013},
which has been considered to be a playground for studying fractionalized magnetic monopoles.

\section*{Acknowledgments}
We thank Y. Okabe for fruitful discussions about MC simulations.
This work was supported by a Grant-in-Aid for Scientific Research on Priority Areas
``Novel States of Matter Induced by Frustration' 
from the Ministry of Education, Culture, Sports, Science and Technology.

\bibliography{reference_DTO.bib}

\begin{thebibliography}{10}

\bibitem{Diep}
H.~T. Diep:  {\em {Frustrated Spin Systems}} (Singapore: World Scientific, Boca
  Raton FL, 2004)
\bibitem{LaCroix}
C. Lacroix, P. Mendels, and F. Mila:  {\em {Introduction to Frustrated
  Magnetism: Materials, Experiments, Theory}} (Springer, New York, 2011)
\bibitem{RamirezNature1999}
A.~P. Ramirez, A. Hayashi, R.~J. Cava, R. Siddharthan, and B.~S. Shastry:
  Nature {\bf 399} (1999) 333.
\bibitem{BramwellScience2001}
S.~T. Bramwell and M.~J.~P. Gingras:  Science {\bf 16} (2001) 1495.
\bibitem{HarrisPRL1997}
M.~J. Harris, S.~T. Bramwell, D.~F. McMorrow, T. Zeiske, and K.~W. Godfrey:
  Phys. Rev. Lett. {\bf 79} (1997) 2554.
\bibitem{BramwellPRL2001}
S.~T. Bramwell, M.~J. Harris, B.~C. den Hertog, M.~J.~P. Gingras, J.~S.
  Gardner, D.~F. McMorrow, A.~R. Wildes, A.~L. Cornelius, J.~D.~M. Champion,
  R.~G. Melko, and T. Fennell:  Phys. Rev. Lett. {\bf 87} (2001) 047205.
\bibitem{MatsuhiraJPCM2000}
K. Matsuhira, Y. Hinatsu, K. Tenya, and T. Sakakibara:  J. Phys.: Condens.
  Matter {\bf 12} (2000) L649.
\bibitem{KadowakiPRB2002}
H. Kadowaki, Y. Ishii, K. Matsuhira, and Y. Hinatsu:  Phys. Rev. B {\bf 65}
  (2002) 144421.
\bibitem{BernalJCP1933}
J. Bernal and R. Fowler:  J. Chem. Phys. {\bf 1} (1933) 515.
\bibitem{RyzhkinJETP2005}
I.~A. Ryzhkin:  JETP {\bf 101} (2005) 481.
\bibitem{CastelnovoNature2008}
C. Castelnovo, R. Moessner, and S.~L. Sondhi:  Nature {\bf 451} (2008) 42.
\bibitem{JaubertNaturePhys2009}
L.~D.~C. Jaubert and P.~C.~W. Holdsworth:  Nat. Phys. {\bf 5} (2009) 258.
\bibitem{Jaubert2011}
L.~D.~C. Jaubert and P.~C.~W. Holdsworth:  J. Phys.: Condens. Matter {\bf 23}
  (2011) 164222.
\bibitem{MatsuhiraJPCM2001}
K. Matsuhira, Y. Hinatsu, and T. Sakakibara:  J. Phys.: Condens. Matter {\bf
  13} (2001) L737.
\bibitem{SnyderPRB2004}
J. Snyder, B.~G. Ueland, J.~S. Slusky, H. Karunadasa, R.~J. Cava, and P.
  Schiffer:  Phys. Rev. B {\bf 69} (2004) 064414.
\bibitem{MatsuhiraJPSJ2011}
K. Matsuhira, C. Paulsen, E. Lhotel, C. Sekine, Z. Hiroi, and S. Takagi:  J.
  Phys. Soc. Jpn. {\bf 80} (2011) 123711.
\bibitem{YaraskavitchPRB2012}
L.~R. Yaraskavitch, H.~M. Revell, S. Meng, K.~A. Ross, H.~M.~L. Noad, H.~A.
  Dabkowska, B.~D. Gaulin, and J.~B. Kycia:  Phys. Rev. B {\bf 85} (2012)
  020410(R).
\bibitem{RevellNaturePhys2012}
H.~M. Revell, L.~R. Yaraskavitch, J.~D. Mason, K.~A. Ross, H.~M.~L. Noad, and
  H.~A. Dabkowska:  Nat. Phys. {\bf 9} (2012) 34.
\bibitem{QuilliamPRB2011}
J.~A. Quilliam, L.~R. Yaraskavitch, H.~A. Dabkowska, B.~D. Gaulin, and J.~B.
  Kycia:  Phys. Rev. B {\bf 83} (2011) 094424.
\bibitem{HertogPRL2000}
B.~C. den Hertog and M.~J.~P. Gingras:  Phys. Rev. Lett. {\bf 84} (2000) 3430.
\bibitem{MelkoJPCM2004}
R.~G. Melko and M.~J.~P. Gingras:  J. Phys.: Condens. Matter {\bf 16} (2004)
  R1277.
\bibitem{RuffPRL2005}
J.~P.~C. Ruff, R.~G. Melko, and M.~J.~P. Gingras:  Phys. Rev. Lett. {\bf 95}
  (2005) 097202.
\bibitem{YavorsPRL2008}
T. Yavors'kii, T. Fennell, M.~J.~P. Gingras, and S.~T. Bramwell:  Phys. Rev.
  Lett. {\bf 101} (2008) 037204.
\bibitem{TabataPRL2006}
Y. Tabata, H. Kadowaki, K. Matsuhira, Z. Hiroi, N. Aso, E. Ressouche, and B.
  Fak:  Phys. Rev. Lett. {\bf 97} (2006) 257205.
\bibitem{PiccoPRB2001}
M. Picco, F. Ricci-Tersenghi, and F. Ritort:  Phys. Rev. B {\bf 63} (2001)
  174412.
\bibitem{SnyderNature2001}
J. Snyder, J.~S. Slusky, R.~J. Cava, and P. Schiffer:  Nature {\bf 413} (2001)
  48.
\bibitem{BovoNatureComm2013}
L. Bovo, J.~A. Bloxsom, D. Prabhakaran, G. Aeppli, and S.~T. Bramwell:  Nat.
  Commun. {\bf 4} (2013) 1535.
\bibitem{Aharoni1998}
A. Aharoni:  J. Appl. Phys. {\bf 83} (1998) 3432.
\bibitem{CastelnovoPRB2011}
C. Castelnovo, R. Moessner, and S.~L. Sondhi:  Phys. Rev. B {\bf 84} (2011)
  144435.
\bibitem{GasparovPRB2012}
V.~A. Gasparov and I. Bo$\breve{\rm z}$ovi$\acute{\rm c}$:  Phys. Rev. B {\bf
  86} (2012) 094523.
\bibitem{MorrisScience2009}
D.~J.~P. Morris, D.~A. Tennant, S.~A. Grigera, B. Klemke, C. Castelnovo, R.
  Moessner, C. Czternasty, M.~M. K. C. R. J.~U. Hoffmann, K. Kiefer, S.
  Gerischer, D. Slobinsky, and R.~S. Perry:  Science {\bf 326} (2009) 411.
\bibitem{Vivtor}
V.~F. Petrenko and R.~W. Whitworth:  {\em {Physics of Ice}} (Oxford University
  Press, New York, 1999)
\bibitem{MatthewsPRB2012}
M.~J. Matthews, C. Castelnovo, R. Moessner, S.~A. Grigera, D. Prabhakaran, and
  P. Schiffer:  Phys. Rev. B {\bf 86} (2012) 214419.
\bibitem{H.TakatsuJPSJL2013}
H. Takatsu, K. Goto, H. Otsuka, R. Higashinaka, K. Matsubayashi, Y. Uwatoko,
  and H. Kadowaki:  J. Phys. Soc. Jpn. {\bf 82} (2013) 073707.
\end{thebibliography}

\end{document}